\documentclass[hyper,a4paper]{JHEP3}

\usepackage{amsmath,amssymb}
\usepackage[numbers,sort&compress]{natbib}
\usepackage{graphicx}
\usepackage{wrapfig}
\usepackage{ucs}
\usepackage{hepunits}
\usepackage[utf8x]{inputenc}

\newcommand{\mpl}{M_{\rm pl}}
\newcommand{\dd}{\ensuremath{\mathrm{d}}}
\renewcommand{\bar}{\overline}
\newcommand{\etanaive}{\eta_{\text{na\"{\i}ve}}}
\newcommand{\eps}{\epsilon}
\title{The everpresent $\eta$-problem: knowledge of all hidden sectors required}
\author{Sjoerd Hardeman$^{a}$, Johannes M. Oberreuter$^{b}$, Gonzalo A. Palma$^{c}$, Koenraad Schalm$^{a}$ and Ted van der Aalst$^{a}$\\
$^{a}$Instituut-Lorentz for Theoretical Physics, Universiteit Leiden \mbox{Niels Bohrweg 2, Leiden, The Netherlands} \\
$^{b}$Instituut voor Theoretische Fysica, Universiteit van Amsterdam \mbox{Science Park 904, Amsterdam, The Netherlands}\\
$^{c}$Physics Department, FCFM, Universidad de Chile \mbox{Blanco Encalada 2008, Santiago, Chile} \\
Email: \email{sjoerd@lorentz.leidenuniv.nl}, \email{j.m.oberreuter@uva.nl}, \email{gpalmaquilod@ing.uchile.cl}, \email{kschalm@lorentz.leidenuniv.nl}, \email{vdaalst@lorentz.leidenuniv.nl}
}

\abstract{We argue that the $\eta$-problem in supergravity inflation cannot be solved without knowledge of the ground state of hidden sectors that are gravitationally coupled to the inflaton. If the hidden sector breaks supersymmetry independently, its fields cannot be stabilized during cosmological evolution of the inflaton. We show that both the subsequent dynamical mixing between sectors as well as the lightest mass of the hidden sector are set by the scale of supersymmetry breaking in the hidden sector. The true cosmological $\eta$-parameter arises from a linear combination of the  lightest mode of the hidden sector with the inflaton. Generically, either the true $\eta$ deviates considerably from the na\"{\i}ve $\eta$ implied by the inflaton sector alone, or one has to consider a multifield model. Only if the lightest mass in the hidden sector is much larger than the inflaton mass and if the inflaton mass is much larger than the scale of hidden sector supersymmetry breaking, is the effect of the hidden sector on the slow-roll dynamics of the inflaton negligible.}

\preprint{ITFA10-44}

\begin{document}
\section{Introduction}
\label{sec:intro}
The construction of realistic models of slow-roll inflation in supergravity is a longstanding puzzle. Supersymmetry can alleviate the finetuning necessary to obtain slow-roll inflation --- if one assumes that the inflaton is a modulus of the supersymmetric ground state --- but cannot solve it completely. This is most clearly seen in the supergravity $\eta$-problem: if the inflaton is a lifted modulus, then its mass in the inflationary background is proportional to the supersymmetry breaking scale. Therefore, the slow-roll parameter $\eta \simeq V''/V$ generically equals unity rather than a small number \cite{Copeland:1994vg}.

We will show here, however, that the $\eta$-problem is more serious than a simple hierarchy problem. In the conventional mode of study, the inflaton sector is always a subsector of the full supergravity theory presumed to describe our Universe. When the inflationary subsector of the supergravity is studied {\em an sich}, tuning a few parameters of the Lagrangian to order $10^{-2}$ will generically solve the problem. We will clarify that this split of the supergravity sector into an inflationary sector and other hidden sectors implicitly makes the assumption that all the other sectors are in a `supersymmetric' ground state: i.e. if the inflaton sector which must break supersymmetry is decoupled, the ground state of the remaining sectors is supersymmetric. If this is not the case, the effect on the $\eta$-parameter or on the inflationary dynamics in general can be large, even if the sypersymmetry breaking scale in the hidden sector is small. Blind truncation in supergravities to the inflaton sector alone, if one does not know whether other sectors preserve supersymmetry, is therefore an inconsistent approach towards slow-roll supergravity inflation. Coupling the truncated sector back in completely spoils the na\"{\i}ve solution found. This result, together with recent qualitatively similar findings for sequestered supergravities (where only the potential has a two-sector structure) \cite{Berg:2010ha}, provides strong evidence that to find true slow-roll inflation in supergravity one needs to know the global ground state of the system. The one obvious class of models  where sector-mixing is not yet considered is the newly discovered manifest embedding of single field inflationary models in supergravity \cite{Kallosh:2010xz}. If these models are also sensitive to hidden sectors, it would arguably certify the necessity of a global analysis for cosmological solutions in supergravity and string theory.

We will obtain our results on two-sector supergravities by an explicit calculation. The gravitational coupling between the hidden and the inflaton sectors is universal, which can be described by a simple $F$-term scalar supergravity theory. As in most discussions on inflationary supergravity theories, we will ignore $D$-terms as one expects its VEV to be zero throughout the early Universe \cite{Lyth:2009zz}. Including $D$-terms (which themselves always need to be accompanied by $F$-terms) only complicates the $F$-term analysis, which is where the $\eta$-problem resides. Furthermore, although true inflationary dynamics ought to be described in a fully kinetic description \cite{Achucarro:2010da}, we can already make our point by simply considering the mass eigenmodes of the system. In a strict slow-roll and slow-turn approximation the mass eigenmodes of the system determine the dynamics of the full system.

Specifically we shall show the following for two-sector supergravities where the sectors are distinguished by independent R-symmetry invariant K\"{a}hler functions:
\begin{itemize}
 \item Given a na\"{\i}ve supergravity solution to the $\eta$-problem, this solution is only consistent if the other sector is in its supersymmetric ground state.
 \item If it is not in its ground state, then the scalar fields of that sector cannot be static but \emph{must} evolve cosmologically as well.
 \item In order for the na\"{\i}ve solution to still control the cosmological evolution these fields must move very slowly. This translates in the requirement that the contribution to the first slow-roll parameter of the hidden sector must be much smaller than the contribution from the na\"{\i}ve inflaton sector, $\eps_{\textrm{hidden}} \ll \eps_{\textrm{na\"{\i}ve}}$.
 \item There are two ways to ensure that $\eps_{\textrm{hidden}}$ is small: Either the supersymmetry breaking scale in the hidden sector is very small or a particular linear combination of first and second derivatives of the generalized K\"ahler function is small.
 \begin{itemize}
  \item In the latter case, one finds that the second slow-roll parameter $\eta_{\textrm{na\"{\i}ve}}$ receives a very large correction $\eta_{\textrm{true}}-\eta_{\textrm{na\"{\i}ve}}\gg \eta_{\textrm{na\"{\i}ve}}$, unless the supersymmetry breaking scale in the hidden sector is small. This returns us to the first case.
  \item In the first case, one finds that the hidden sector always contains a light mode, because in a supersymmetry breaking (almost) stabilized supergravity sector there is always a mode that scales with the scale of supersymmetry breaking. This light mode will overrule the na\"{\i}ve single field inflationary dynamics.
 \end{itemize}
\end{itemize}
Thus for \emph{any} nonzero supersymmetry breaking scale in the hidden sector --- even when this scale is very small --- the true mass eigenmodes of the system are linear combinations of the hidden sector fields and the inflaton sector fields. We compute these eigenmodes. By assumption, the true value of the slow-roll parameter $\eta$ is the smallest of these eigenmodes. Depending on the values of the supersymmetry breaking scale and the na\"{\i}ve lowest mass eigenstate in the hidden sector, we find that
\begin{enumerate}
\item The new set of mass eigenmodes can have closely spaced eigenvalues, and thus the initial assumption of single field inflation is incorrect. Then a full multifield re-analysis is required.
\item The relative change of the value of $\eta$ from the na\"{\i}ve to the true solution can be quantified and shows that for a supersymmetry breaking hidden sector, the na\"{\i}ve model is only reliable if the na\"{\i}ve lowest mass eigenstate in the hidden sector is much larger than the square of the scale of hidden sector supersymmetry breaking divided by the inflaton mass. This effectively excludes all models where the hidden sector has (nearly) massless modes.
\item The smallest eigenmode can be dominantly determined by the hidden sector, and thus the initial assumption that the cosmological dynamics is constrained to the inflaton sector is incorrect. Again a full multifield re-analysis is required.
\end{enumerate}
One concludes that in general one needs to know/assume the ground states and the lowest mass eigenstates of {\em all} the hidden sectors to reliably find a slow-roll inflationary supergravity.

The structure of our paper is the following. Section \ref{sec:sugra} reviews some definitions in supergravity and explains how sectors are coupled in supergravity. This leads directly to the first result that in a stabilized supergravity sector there always is a mode that scales with the scale of supersymmetry breaking. In section \ref{sec:inflation} we discuss the $\eta$-problem in a single sector theory and then consider the effect of a hidden sector qualitatively and quantitatively. The quantitative result is analysed in section \ref{sec:diagram} both in terms of effective parameters and direct supergravity parameters. As a notable example of our result, we show that if the hidden sector is the Standard Model, where  its supersymmetry breaking is not caused by the inflaton sector but otherwise, spoils the na\"{\i}ve slow-roll solution in the putative inflaton sector. The paper is supplemented with two appendices in which some of the longer formulae are given.

\section{A stabilized sector in a supergravity two-sector system}
\label{sec:sugra}
We shall start by recalling how two sectors are gravitationally coupled in supergravity. Although this coupling is universal, the definition differs from regular gravity in an important way: the superpotentials multiply rather than add.

We will then consider one of the two sectors to be a stable hidden sector. We show that a light mode develops, which indicates that the hidden sector obtains a flat direction and is not stable any more. This extends the result of \cite{GomezReino:2006dk}, in which it is shown that non-supersymmetric Minkowski minima always develop at least one light mass mode, to de Sitter and Anti-de Sitter vacua.

\subsection{The supergravity action}
The action for the scalar sector of $\mathcal{N}\! = \! 1$ supergravity is
\begin{equation}\label{eq:fullaction}
S = \mpl^{2} \int \dd^{4}x\sqrt{g}\left[\frac{1}{2}R -
g^{\mu\nu}G_{\alpha\bar{\beta}}\partial_{\mu}X^{\alpha}\partial_{\nu}\bar{X}^{\bar{\beta}}
- V \mpl^2\right] \;,
\end{equation}
in which $G_{\alpha\bar{\beta}}$ is the field space metric and $g_{\mu\nu}$ is the spacetime metric with associated Riemann scalar $R$. The Greek indices run over all fields $\{\alpha,\bar{\beta}\}$ or over spacetime coordinates $\{\mu,\nu\}$. For calculational convenience we have defined the scalar fields $X$ and functions $V$, $K$ and $W$ to be dimensionless. The ($F$-term) potential $V$ of the scalar sector is defined as
\begin{equation}\label{eq:scalarpotential}
V = e^{G}\left(G_{\alpha}G^{\alpha} - 3\right) \;.
\end{equation}
Through $G_{\alpha} = \partial_{\alpha} G$, $G_{\alpha\bar{\beta}}=\partial_\alpha\partial_{\bar{\beta}}G$, the action \eqref{eq:fullaction} is completely specified by the real K\"{a}hler function $G(X,\bar{X})$, which is related to global supersymmetry quantities through
\begin{equation}
G(X,\bar{X}) = K(X,\bar{X}) + \log\left(W(X) \right) + \log\left( \bar{W}(\bar{X}) \right)
\end{equation}
in terms of the real K\"{a}hler potential $K(X,\bar{X})$ and the holomorphic (dimensionless) superpotential $W(X)$.\footnote{Note that this definition requires $W \neq 0$. For $W=0$ a K\"ahler function $G$ cannot be defined. In this paper we will assume that $W\neq0$.} The definition for $G$ is convenient as it is invariant under K\"{a}hler transformations, i.e. it is invariant under the simultaneous transformation of $K(X,\bar{X}) \rightarrow K(X,\bar{X}) + f(X) +\bar{f}(\bar{X})$ and $W(X) \rightarrow e^{-f(X)}W(X)$ for an arbitrary holomorphic function $f(X)$.

\subsection{Canonical coupling}
\label{sec:canonicalcoupling}
To describe a two-sector system we consider a class of minimally coupled scenarios \cite{Cremmer:1982vy,Binetruy:1984wy,BenDayan:2008dv}
\begin{equation}\label{eq:twosectorG}
G(\phi,\bar{\phi},q,\bar{q}) = G^{(1)}(\phi,\bar{\phi}) + G^{(2)}(q,\bar{q}) \;,
\end{equation}
with $\phi, q$ denoting the fields in the two sectors respectively. In the following, we will take the indices $\{i,\bar{\jmath}\}$ to run over the $\phi$-fields, while $\{a,\bar{b}\}$ denote the fields in the $q$-sector. Later in the paper we will take the $\phi$-fields to drive inflation, while the $q$-fields reside in another sector which is na\"{\i}vely assumed not to take part in the inflationary dynamics and is hence called the hidden sector. This split of the K\"{a}hler function $G(\phi,\bar{\phi},q,\bar{q})$ (\ref{eq:twosectorG}) is invariant under K\"{a}hler transformations in each sector separately \cite{Choi:2004sx,deAlwis:2005tf,deAlwis:2005tg,Achucarro:2007qa,Achucarro:2008sy} and thus defines a sensible way of splitting up the action in multiple sectors. Amongst other properties, this split guarantees that a BPS solution in one particular sector is a BPS solution of the full theory. In terms of $K$ and $W$, this definition has a conventional separation of the K\"ahler potential, but the superpotentials in each sector combine multiplicatively rather than add
\begin{equation}\label{eq:defGinKW}
K(\phi,\bar{\phi},q,\bar{q}) + \log |W(\phi,q)|^{2} = K^{(1)}(\phi,\bar{\phi}) + K^{(2)}(q,\bar{q}) + \log|W^{(1)}(\phi)W^{(2)}(q)|^{2}\;.
\end{equation}

Let us illustrate the importance of this multiplicative superpotential in the situation in which the hidden sector resides in a supersymmetric vacuum, i.e. $\partial_a V(q_0)=0$ and $\partial_a G^{(2)}(q_0)=0$. We write the superpotential of the hidden sector as $W^{(2)}(q)=W^{(2)}_0+W^{(2)}_{\textrm{global}}(q-q_0)$. The second term in this expression is what determines the potential for fluctuations around the minimum of the hidden sector, while the first constant term is just an overall contribution and hence not interesting for the internal hidden sector dynamics at energies much less than the Planck scale. However, for the gravitational dynamics and the remaining $\phi$-sector this `vacuum energy contribution' $W^{(2)}_0$ is of crucial importance as it sets the scale of the potential
\begin{equation}
V=e^{K^{(2)}}|W^{(2)}_0|^2 e^{G^{(1)}}\left(G^{(1)}_{i}G^{(1)i}-3\right) \;,
\end{equation}
which is evaluated at $q=q_0$ such that all terms depending on $W^{(2)}_{\textrm{global}}$ vanish. The normal practice of setting $W^{(2)}_0$ to zero as an overall contribution to the hidden sector is neglecting the fact that gravity also feels the constant part of the potential energy, as opposed to field theory. The inflationary sector feels the presence of the hidden sector through this coupling and as such it may be more intuitive to regard $W^{(2)}_0$ to contain information about the inflationary sector rather than the hidden sector. Making a similar split in $W^{(1)}$, the constant part $W^{(1)}_0$ is the overall contribution to the hidden sector due to the inflaton sector.

The multiplicative superpotential also means that the zero-gravity limit to a global supersymmetry is more subtle than just taking $\mpl \to \infty$, as is usually done \cite{Achucarro:2011yc}. One must first determine a ground state which sets $W^{(1)}_0$ and $W^{(2)}_0$, and then send both $W^{(1)}_0\to 0$ and $W^{(2)}_0\to 0$ in such a way that the combinations $W^{(1)}_0\mpl$ and $W^{(2)}_0\mpl$ remain constant. Instating the canonical dimensions for the fields and the K\"{a}hler potential and rescaling the couplings such that $W^{(2)}_{\textrm{eff}} = W^{(1)}_0W^{(2)}_{\textrm{global}}$ and  $W^{(1)}_{\textrm{eff}} = W^{(2)}_0W^{(1)}_{\textrm{global}}$ scale as $\mpl^{-3}$, the total superpotential
\begin{equation}\label{eq:totalsplitpotential}
W=W^{(1)}_0W^{(2)}_0+W^{(1)}_0W^{(2)}_{\textrm{global}}+W^{(2)}_0W^{(1)}_{\textrm{global}}+W^{(1)}_{\textrm{global}}W^{(2)}_{\textrm{global}}
\;,
\end{equation}
then consists of a constant term which scales as $W^{(1)}_0W^{(2)}_0 \sim \mpl^{-2}$, crossterms which scale as $W^{(1)}_0W^{(2)}_{\textrm{global}}+W^{(2)}_0W^{(1)}_{\textrm{global}} \sim \mpl^{-3}$ and a multiplicative term which scales as  $W^{(1)}_{\textrm{global}}W^{(2)}_{\textrm{global}} \sim \mpl^{-4}$. Considering the dimensionful superpotential this results in an overall infinite contribution, a finite sum of two terms and a vanishing product. In this decoupling limit one recovers the two independent global supersymmetry sectors with the na\"{\i}ve additive behavior in both the superpotential and the K\"{a}hler potential,
\begin{align}\label{eq:defglobalsusy}
K(\phi,\bar{\phi},q,\bar{q}) &= K^{(1)}(\phi,\bar{\phi}) + K^{(2)}(q,\bar{q}) \;, \nonumber \\
W(\phi,q) &= W^{(1)}_{\textrm{eff}}(\phi) + W^{(2)}_{\textrm{eff}}(q) \;.
\end{align}
However, one cannot use this split \eqref{eq:defglobalsusy} and couple gravity back in \cite{Davis:2008sa}. As explained, in supergravity the definition \eqref{eq:defglobalsusy} is not invariant under K\"{a}hler transformations in each sector separately and is valid only in a specific K\"{a}hler frame or, say, gauge dependent \cite{Achucarro:2007qa}. Another way to understand the result is to realise that the definition \eqref{eq:defglobalsusy} does not lead to a K\"ahler metric and mass matrix that can be made block diagonal in the same basis \cite{Achucarro:2008sy}, and thus there is no sense of `independent' sectors.

Insisting on the separate K\"ahler invariance of (\ref{eq:twosectorG}), the two-sector action \eqref{eq:fullaction} reads
\begin{align}\label{eq:splitaction}
S &= \mpl^2\int \dd^{4}x\sqrt{g}\left[\frac{1}{2}R -
g^{\mu\nu}(G^{(1)}_{i\bar{\jmath}}\partial_{\mu}\phi^i\partial_{\nu}\bar{\phi}^{\bar{\jmath}}
+
G^{(2)}_{a\bar{b}}\partial_{\mu}q^a\partial_{\nu}\bar{q}^{\bar{b}})
- V\mpl^2\right],
\end{align}
with
\begin{align}
V(\phi,\bar{\phi},q,\bar{q})&=e^{G^{(1)} +
G^{(2)}}\left(G^{(1)}_{i}G^{(1)i} + G^{(2)}_{a}G^{(2)a} - 3\right)
\;.
\end{align}
We will allow ourselves to drop the sector label from $G$ in the remainder, since $G^{(1)}_{\phi} = G_{\phi}$ and similarly for $q$. For a short overview of relevant conventions and identities in supergravity, we refer the reader to appendix \ref{app:sugrarelations}.

\subsection{Zero mass mode for a stabilized sector}\label{sec:zeromode}
Anticipating the situation for an inflationary scenario we will analyse the mass spectrum of a stabilized $q$-sector in a de Sitter background. For Minkowski spaces it is known that the lightest mass in a stabilized sector scales with the supersymmetry breaking VEV $G_{a}$ \cite{GomezReino:2006dk}. Here we extend the analysis to de Sitter vacua as the zeroth order approximation of slow-roll inflation. Already in this zeroth order approach we will show that a similar light mode develops in the stabilized sector. Throughout this discussion we assume that the potential $V$ is kept positive by the presence of the `inflationary' sector. In the next section we show that this result can be translated directly into an inflationary setting, where this light mode will affect the slow-roll dynamics.

Given that we insist the $q$-sector to be stabilized, we have $\partial_a V=0$. In terms of the K\"ahler function $G(\phi,\bar{\phi},q,\bar{q})$ this means
\begin{equation}
 (\nabla_aG_b)G^b=-G_a(1+e^{-G}V) \;.
\end{equation}
If the $q$-ground state breaks supersymmetry, i.e. $G_a\neq 0$, we may rewrite it in terms of the supersymmetry breaking direction $f_a=G_a/\sqrt{G^bG_b}$,
\begin{equation}\label{eq:Vq=0conditionmultiplefields}
 (\nabla_aG_b)f^b=-f_a(1+e^{-G}V) \;.
\end{equation}
For simplicity we will assume that the $q$-sector consists of only a single complex scalar field $q$, in which case we may write this equation as
\begin{equation}\label{eq:Vq=0condition}
 \nabla_qG_q=-G_{q\bar{q}}(1+e^{-G}V)\widehat{G}_q^{2} \;.
\end{equation}
A hat $\widehat{z}$ on a complex number denotes the `phase'-part of the number, $z=|z|\widehat{z}=|z|e^{i\;\textrm{arg}(z)}$. As such $\widehat{G}_q=\sqrt{G^{q\bar{q}}}f_q$. Note that in an arbitrary supersymmetric configuration $G_a=0$ there are no restrictions on $\nabla_a G_b$, but on a supersymmetry broken configuration this is no longer true. Were one to turn on supersymmetry breaking, one would first have to reach a surface in parameter space where this restriction can be imposed at the onset of supersymmetry breaking.

We will now compute the mass spectrum for the two modes of the complex scalar field $q$, at the hypersurface defined by \eqref{eq:Vq=0condition}. The mass modes are given by the eigenvalues of the matrix
\begin{equation}
M^2=\begin{pmatrix}V^q_{\phantom{q}q}&V^q_{\phantom{q}\bar{q}}\\V^{\bar{q}}_{\phantom{\bar{q}}q}&V^{\bar{q}}_{\phantom{\bar{q}}\bar{q}}\end{pmatrix},
\end{equation}
which in our case means
\begin{equation}\label{eq:massmodes}
m_q^{\pm}=\left(V^q_{\phantom{q}q}\pm|V^q_{\phantom{q}\bar{q}}|\right)=G^{q\bar{q}}\left(V_{q\bar{q}}\pm|V_{qq}|\right).
\end{equation}
Expanding the second derivatives of the potential (cf.~appendix \ref{app:masslessmode}) to first order in $|G_q|$, these eigenvalues are
\begin{align}
m_q^-&=e^G G^{q\bar{q}}\textrm{Re}\big\{(\nabla_q\nabla_qG_q)\widehat{G^q}^3\big\}|G^q|+\mathcal{O}(|G_q|^2)\;,\label{eq:zeromassmode}\\
m_q^+&=e^G\left[2(2+e^{-G}V)(1+e^{-G}V)-G^{q\bar{q}}\textrm{Re}\big\{(\nabla_q\nabla_qG_q)\widehat{G^q}^3\big\}|G^q|\right]+\mathcal{O}(|G_q|^2)\;.\label{eq:heavymassmode}
\end{align}
We see from \eqref{eq:zeromassmode} that in the limit of vanishing supersymmetry breaking the lightest mass mode becomes massless, just as in the case of Minkowski space \cite{GomezReino:2006dk}.\footnote{The result can also be extended to hold for anti-de Sitter vacua. However, for $-2<e^{-G}V<-1$, also a tachyonic mode develops.} It is important to note that this result depends crucially on taking the limit $G_{q}$ to zero in the supersymmetry breaking direction. When supersymmetry is restored and both $G_{q}=0$ and $G_{\bar{q}}=0$, the phases of these vectors have no meaning. In fact, we see that then a new degree of freedom arises: $\nabla_qG_q$ becomes unrestricted which allows one to choose the masses freely.

The geometrical picture is that there is a whole plane of supersymmetric solutions where arbitrary masses are allowed. However, when supersymmetry is broken, the supersymmetry breaking direction has to align with its complex conjugate fixing one point on this plane where supersymmetry can be broken. In this point, the lightest mode becomes massless.

\section{Two-sector inflation in supergravity}
\label{sec:inflation}
Generally, when inflation is described in supergravity, realistic matter resides in a hidden sector.\footnote{The supersymmetric partners of the Standard Model are not good inflaton candidates, as these partners are charged under the Standard Model gauge group and gauge fields taking part in inflation would lead to topological defects \cite[eg.][]{Kibble:1976sj,Jeannerot:2003qv}. The exception could be a gravitationally non-minimally coupled Higgs field \cite[eg.][]{Bezrukov:2007ep,Germani:2010gm}.} Supergravities descending from string theory often have additional hidden sectors as well. These sectors are always gravitationally coupled. In the previous section we have seen that for de Sitter vacua the hidden sector develops a light direction. In this section we will consider how this light mode of the hidden sector can affect the na\"{\i}ve dynamics of the inflationary sector. We will show that despite the weakness of gravity, these effects can be large. Realistic slow-roll inflation is characterized by small numbers, the slow-roll parameters $\eps$ and $\eta$, and even small absolute changes to these numbers can be of the order of 100\% in relative terms.

We will first briefly review the $\eta$-problem in the context of single field inflation in supergravity. Then we will explain what effects are to be expected when including an additional (hidden) sector. The section ends with calculating the relevant objects to determine the true dynamics of the full system.

\subsection{Inflation and the $\eta$-problem in supergravity}\label{sec:singlefield}
In single scalar field models of inflation the spectrum of density perturbations is characterized by the two slow-roll parameters $\eps$ and $\eta$. To ensure that this spectrum matches the observed near scale invariance, both $\eps\ll 1$ and $\eta \ll 1$. Inflationary supergravity in its simplest form consists of a single complex scalar field, the inflaton, whose potential is generated by $F$-terms \eqref{eq:scalarpotential}. The definition of $\eta$ may be phrased as the lightest direction of the mass matrix in units of the Hubble rate $3 H^{2} = V$, i.e. $\eta$ is the smallest eigenvalue of the matrix
\cite{Burgess:2004kv}
\begin{equation}\label{eq:etanaive}
\widetilde{N}^{I}_{\phantom{I}J} = \frac{1}{V}
 \begin{pmatrix}
  \nabla^{i}\nabla_{j} V&  \nabla^{i}\nabla_{\bar{\jmath}} V \\
  \nabla^{\bar{\imath}}\nabla_{j} V &  \nabla^{\bar{\imath}}\nabla_{\bar{\jmath}} V
 \end{pmatrix}\;,
\end{equation}
where the tilde on $\widetilde{N}$ indicates that this value of $\eta$ is defined with respect to the inflaton sector only and $I\in\{i,\bar{\imath}\}$, $J\in\{j,\bar{\jmath}\}$, respectively.\footnote{A careful definition based on the kinetic behaviour of the inflaton field is done in \cite{GrootNibbelink:2000vx,GrootNibbelink:2001qt}. In the slow-roll, slow-turn limit, it reduces to the definition of  $\eta$ given here.} From the second $\phi$-derivative of $V$,
\begin{equation}
V_{i\bar{\jmath}}=G_{i\bar{\jmath}}V+G_{i} V_{\bar{\jmath}}+G_{\bar{\jmath}}V_{i}-G_{i} G_{\bar{\jmath}}V +e^{G}\left[R_{i\bar{\jmath}k\bar{l}}G^{k}G^{\bar{l}}+G^{k\bar{l}}\nabla_{i} G_{k}\nabla_{\bar{\jmath}}G_{\bar{l}}+G_{i\bar{\jmath}}\right]\;,
\end{equation}
we see that a natural value for $\eta$ is $V^i_{\phantom{i}j}/V\sim\nabla^iG_j \sim 1$ is unity. Therefore, we must tune $G_{i}$, $\nabla_{i} G_{j}$ and $R_{i\bar{\jmath}k\bar{l}}$ so that $V^i_{\phantom{i}j}=\mathcal{O}(10^{-3})V$. The necessity of this tuning is known as the $\eta$-problem.

As shown in \cite{Covi:2008cn}, successful inflation is achievable if one tunes the K\"{a}hler function $G$ such that
\begin{equation}
R_{i\bar{\jmath}k\bar{l}}f^{i}f^{\bar{\jmath}}f^{k}f^{\bar{l}} \lesssim \frac{2}{3} \frac{1}{1+\gamma}\;,
\end{equation}
where $ \gamma= e^{-G}V/3$ is inversely proportional to an overall mass scale $m_{3/2}=e^{G/2}$, which is related to the gravitino mass and $R_{i\bar{\jmath}k\bar{l}}$ is the Riemann tensor of the inflaton sector. As $f^{i}f_{i} = 1$ the above equation defines the normalised sectional curvature along the direction of supersymmetry breaking. The constraint becomes stronger as $\gamma \gg 1$, thus as $H \gg m_{3/2}$. When the bound is met, one can always tune $\eta$ to be small by tuning $G_{i}$, $\nabla_{i}G_{j}$ and $R_{i\bar{\jmath}k\bar{l}}$.

Finding a suitably tuned supergravity potential from a (UV-complete) string theoretical set-up has proven to be incredibly difficult \cite{Covi:2008ea,Covi:2008zu}, but possible \cite{Baumann:2007np,Kachru:2003aw,Giddings:2001yu}. Currently, in models with correctly tuned slow-roll parameters it is typically assumed that the `hidden sector' does not affect the finetuning of parameters. The subject of this paper is to examine whether such an assumption is justified and hence how relevant tuned models are that only consider the inflationary sector.

\subsection{Stability of the hidden sector during inflation}\label{sec:nonstablehiddensector}
Having reviewed the $\eta$-problem in single sector supergravity theories, we will now consider if and how the fields in the hidden sector can affect the inflationary evolution. From the diagonalisation of the kinetic terms in \eqref{eq:fullaction} the distinction between $\phi$-fields and $q$-fields is explicit, leading naturally to an inflationary and a hidden sector. We will again assume these sectors to both consist of only one complex scalar field, $\phi$ and $q$ respectively. The argument we shall present can already be made in a two-field system. It carries through to multifield models because the field $\phi$ is viewed as the inflaton in an effective single field inflationary model, while the field $q$ can be seen as the lightest mode in the hidden sector. Following the usual practice \cite[and references therein]{Baumann:2009ni,McAllister:2007bg}, we assume that inflation is solved by tuning the inflationary sector only, including obtaining satisfactory values for the slow-roll parameters from a phenomenological viewpoint. As a result all data in the inflationary sector are fixed and known. Contrarily, the hidden sector is left unspecified and the restrictions we find on it are a function of model specific parameters of the inflaton sector only.

To ensure that the hidden sector does not take part in the inflationary dynamics, one generally assumes that the fields in the hidden sector are stabilized in a ground state at a constant field value $q=q_0$ throughout inflation
\begin{equation}\label{eq:stabilizationcondition}
\left.\partial_q V\right|_{q_0}=0
\end{equation}
and, hence, are not dynamical. Clearly this is true if $G_q=0$, i.e. when the ground state of the hidden sector preserves supersymmetry. As was shown in detail in \cite{deAlwis:2005tf,deAlwis:2005tg,Achucarro:2007qa,Achucarro:2008sy,Achucarro:2008fk,Achucarro:2010jv,Kallosh:2010xz,Kallosh:2010ug}, when $G_q=0$ the ground state of the hidden sector decouples gravitationally from the inflationary sector and the inflationary sector truly determines the inflationary evolution without any contributions from the hidden sector.

The case we examine here is when supersymmetry is broken in the hidden sector, $G_q \neq 0$. The first thing to note is that the stability assumption (\ref{eq:stabilizationcondition}) cannot be met anymore. In supergravity the position $q=q_0$ of the minimum of the potential is given by
\begin{equation}
V_q = G_q V (\phi,\bar{\phi},q,\bar{q}) + e^{G(\phi,\bar{\phi},q,\bar{q})}\left( (\nabla_q G_{q}) G^q + G_q\right) = 0 \;,
\end{equation}
which shows that for $G_q \neq 0$ the ground state $q_0$ depends on the inflaton field $\phi$, through $V(\phi,\bar{\phi},q,\bar{q})$ and $G(\phi,\bar{\phi},q,\bar{q})$. In the situation of unbroken supersymmetry, $G_q=0$, all $\phi$-dependence drops out, but for $G_q\neq 0$ we see that it is impossible to keep the position of the minimum constant during inflation. As the inflaton $\phi$ rolls down the inflaton direction, the `stabilized' hidden scalar $q$ will change its value. It is clear that the assumption of a vanishing $V_q=0$ for all $q$ is incompatible with $G_q\neq 0$ and we should therefore abandon it. This in turn means that the hidden sector field $q$ must be dynamical, through its equation of motion. Since we still want to identify the field $\phi$ as the inflaton in the sense that it drives the cosmological dynamics, we have to assume that $q$ moves very little. We must therefore also assume a slow-roll, slow-turn approximation to the solution of the $q$ equation of motion
\begin{equation}
 \dot{q}=\frac{G^{q\bar{q}}V_{\bar{q}}}{3H}\;.
\end{equation}
The statement that the cosmological dynamics is driven by the $\phi$-sector means that $\|\dot{q}\|\ll\|\dot{\phi}\|$, where $\|\dot{q}\| \equiv \sqrt{G_{q\bar{q}}\dot{q}\dot{\bar{q}}}$, etc. Through both slow-roll equations of motion this equates to $\|V_q\|\ll \|V_\phi\|$ or $\epsilon_q\ll \epsilon_\phi$,

As the hidden sector has now become dynamical, we have to treat the system as a multifield inflationary model. Since it is impossible to diagonalise the K\"{a}hler transformations and mass matrix simultaneously, the fields will mix in the case of a hidden sector with broken supersymmetry \cite{Achucarro:2007qa}. In the next section we will study the consequences of this mixing by explicitly diagonalising the mass matrix of the full two-field system. From the result we shall find three possible effects on the inflationary dynamics.

First, the lightest masses of fields from the different sectors can be too close together. It is obvious that one cannot consider an effective single field model if this is the case, since for the dynamics to be independent of initial conditions, the lightest field needs to be much lighter than the other fields.  When the masses of the two fields are similar, both of them contribute to the dynamics, resulting into a multifield rather than a single field inflationary scenario. As is known from the literature, a multifield inflationary model will produce effects such as isocurvature modes \cite[eg.][]{Mollerach:1989hu,Polarski:1994rz,Linde:1996gt,Hwang:2001fb,Wands:2002bn,DiMarco:2002eb,vanTent:2003mn,Byrnes:2006fr,Wands:2007bd,Lalak:2007vi,Malik:2008im,Langlois:2008vk,Langlois:2008mn,Langlois:2008wt,Langlois:2008qf}, features in the power spectrum \cite[eg.][]{Rigopoulos:2005xx,Peterson:2010np,Cremonini:2010ua,Achucarro:2010da} and non-Gaussianities \cite[isocurvature models and eg.][]{Verde:1999ij,Bartolo:2001cw,Bernardeau:2002jy,Seery:2005gb,Rigopoulos:2005ae,Rigopoulos:2005us,Vernizzi:2006ve,Tolley:2009fg,Chen:2009we,Chen:2009zp}, pointing to a qualitatively different model.

Second, a change of the true value of $\eta$ can occur. We have assumed the inflaton sector to be tuned in such a way that it agrees with observed values for the slow-roll parameters. If the effects of the hidden sector on the total dynamics are such that $\eta$ will change significantly, the initial na\"{\i}ve tuning would be of no meaning and one would have to start the tuning process all over again after the hidden sector has been added. Again we note that there is no contribution in the case of unbroken supersymmetry in the hidden sector, since we shall show that the contribution to $\eta$ from the hidden sector is mostly determined by the cross terms in the mass matrix,
\begin{equation}
 V_{\phi q}=G_{\phi} V_{q}+G_{q}V_{\phi}-G_{\phi} G_{q}V\;,
\end{equation}
which vanish when $G_q=0$.

Third, a complete change of the sector that determines $\eta$ is possible. It is possible that the eventual $\eta$-parameter is still within the limits of its na\"{\i}ve tuned value, satisfying the second bound, but instead it is determined by the hidden sector rather than the inflationary sector. Any initial control obtained by tuning the inflationary sector is superseded by the sheer coincidental configuration of the hidden sector.

\subsection{The mass matrix of a two-sector system}
To investigate when effects from the hidden sector are to be expected, we need to calculate the eigenvalues of the mass matrix of the full two-field system. Since we assume the inflationary evolution to be in the slow-roll, slow-turn regime, the dynamics is completely potential energy dominated. The mass matrix of the full two-field system determines which directions are stable or steep, as characterised by the eigenvalues of this matrix. Normalizing by $1/V$ to obtain the value of $\eta$ directly, the matrix we want to diagonalise is the $4\times4$-matrix
\begin{equation}\label{eq:massmatrix}
N^A_{\phantom{A}B}=\frac{1}{V}
 \begin{pmatrix}
  \nabla^{\alpha}\nabla_{\beta}V&\nabla^{\alpha}\nabla_{\bar{\beta}}V\\
  \nabla^{\bar{\alpha}}\nabla_{\beta}V&\nabla^{\bar{\alpha}}\nabla_{\bar{\beta}}V
\end{pmatrix},
\end{equation}
where
$A\in\{\alpha,\bar{\alpha}\}$ and $B\in\{\beta,\bar{\beta}\}$ run over both fields $\phi$ and $q$ and their complex conjugates. Equation \eqref{eq:massmatrix} is to be evaluated at a point near $q_0=q_0(\phi_0)$, where $q_0$ is such that $\partial_q V(q_0)=0$, with $\phi_0$ indicating the beginning of inflation. As is clear from the discussion of section \ref{sec:nonstablehiddensector} we cannot truly expect the hidden sector to be stabilized throughout the inflationary evolution. Nevertheless we may consider $\partial_q V(q_0) =0$ at a certain point $q_0=q_0(\phi_0)$, with $\|\partial_q V\| \ll \|\partial_\phi V\| $ around $q_0$ in accordance with the restriction $\eps_q \ll \eps_\phi$.

The mass matrix is Hermitian and, considering again a two-field system, can be put in the form
\begin{equation}
N^A_{\phantom{A}B}=\frac{1}{V}
\begin{pmatrix}
  \nabla^{\phi}V_{\phi}&\nabla^{\phi}V_{\bar{\phi}}&\nabla^{\phi}V_{q}&\nabla^{\phi}V_{\bar{q}}\\
  \nabla^{\bar{\phi}}V_{\phi}&\nabla^{\bar{\phi}}V_{\bar{\phi}}&\nabla^{\bar{\phi}}V_{q}&\nabla^{\bar{\phi}}V_{\bar{q}}\\
  \nabla^{q}V_{\phi}&\nabla^{q}V_{\bar{\phi}}&\nabla^{q}V_{q}&\nabla^{q}V_{\bar{q}}\\
  \nabla^{\bar{q}}V_{\phi}&\nabla^{\bar{q}}V_{\bar{\phi}}&\nabla^{\bar{q}}V_{q}&\nabla^{\bar{q}}V_{\bar{q}}
\end{pmatrix},
\end{equation}
by a coordinate transformation. Diagonalising the full matrix in general is involved. Therefore, we adopt the strategy to diagonalise the two sectors separately and then pick the lightest modes only. The first  step yields
\begin{equation}
N^A_{\phantom{A}B}=\begin{pmatrix}\frac{1}{V}(V^{\phi}_{\phantom{\phi}\phi}-|V^\phi_{\phantom{\phi}\bar{\phi}}|)& 0&A_{11}&A_{12}\\
  0&\frac{1}{V}(V^{\phi}_{\phantom{\phi}\phi}+|V^\phi_{\phantom{\phi}\bar{\phi}}|)&A_{21}&A_{22}\\
  \bar{A}_{11}&\bar{A}_{21}&\frac{1}{V}(V^{q}_{\phantom{q}q}-|V^q_{\phantom{q}\bar{q}}|)& 0\\
  \bar{A}_{12}&\bar{A}_{22}& 0&\frac{1}{V}(V^{q}_{\phantom{q}q}+|V^q_{\phantom{q}\bar{q}}|)\end{pmatrix},
\end{equation}
with
\begin{equation}
A= \frac{1}{2V}
  \begin{pmatrix}-\widehat{V_{\bar{\phi}\bar{\phi}}}&\widehat{V_{\bar{\phi}\bar{\phi}}}\\1&1\end{pmatrix}^{-1}
  \begin{pmatrix}V^\phi_{\phantom{\phi}q}&V^\phi_{\phantom{\phi}\bar{q}}\\V^{\bar{\phi}}_{\phantom{\bar{\phi}}q}&V^{\bar{\phi}}_{\phantom{\bar{\phi}}\bar{q}}\end{pmatrix}
  \begin{pmatrix}-\widehat{V_{\bar{q}\bar{q}}}&\widehat{V_{\bar{q}\bar{q}}}\\1&1\end{pmatrix} \;.
\end{equation}
Here, the first matrix is the inverse of the similarity transformation of the $\phi$-sector and the last matrix diagonalises the $q$-sector.

In general the eigenmodes in the individual sectors will be different, one always being smaller than the other. Dynamically the most relevant direction is the lightest mode of each sector, but by restricting to these light directions, one assumes a hierarchy already within the sectors. For the inflationary sector this is phenomenologically justified if we assume that inflation is described by a single field, where we know that $V^\phi_{\phantom{\phi}\phi}$ and $V^\phi_{\phantom{\phi}\bar{\phi}}$ combine such that a light mode appears with mass $\eta V$, much lighter than the other mass modes. For the hidden sector we will simply assume that a large enough hierarchy between mass modes exists. This will simplify matters without weakening our result. By including only the lightest mode of the hidden sector, we can already show that the true dynamics is in many cases not correctly described by the na\"{\i}ve inflaton sector. Our case would only be more strongly supported if we would include the heavy mode of the hidden sector, but this is technically more involved. Projecting on the light directions, we get a submatrix of light mass modes
\begin{equation}
  N_{\mathrm{light}}=\begin{pmatrix}\lambda_\phi&A_{11}\\\bar{A}_{11}&\lambda_q\end{pmatrix},
\end{equation}
with
\begin{align}
\label{eq:lambdaphi}
  \lambda_\phi&=\frac{1}{V}\left(V^\phi_{\phantom{\phi}\phi}-|V^\phi_{\phantom{\phi}\bar{\phi}}|\right)=\frac{G^{\phi\bar{\phi}}}{V}(V_{\phi\bar{\phi}}-|V_{\phi\phi}|)\;,\\
  \label{eq:lambdaq}
\lambda_q&=\frac{1}{V}\left(V^q_{\phantom{q}q}-|V^q_{\phantom{q}\bar{q}}|\right)=\frac{G^{q\bar{q}}}{V}(V_{q\bar{q}}-|V_{qq}|)\;,\\
\label{eq:Amix}
A_{11}&=\frac{G^{\phi\bar{\phi}}}{2V} \left(
\widehat{V_{\bar{q}\bar{q}}}\widehat{V_{\phi\phi}}V_{\bar{\phi}q}-
\widehat{V_{\bar{q}\bar{q}}} V_{\phi q}+ V_{\phi \bar{q}} -
\widehat{V_{\phi \phi}} V_{\bar{\phi}\bar{q}} \right)\;.
\end{align}
The eigenvalues of this two-field system are given by
\begin{equation}\label{eq:neweigenvalues}
\mu_{\pm}=\frac{1}{2}\left(\lambda_\phi+\lambda_q\right)\pm\frac{1}{2}\sqrt{\left(\lambda_q-\lambda_\phi\right)^2+4|A_{11}|^2}\;.
\end{equation}
Since $\mu_-<\mu_+$ the second slow-roll parameter for the full system is given by $\eta=\mu_-$.

\section{Dynamics due to the hidden sector}\label{sec:diagram}
In slow-roll and slow-turn approximation, the mass modes $\mu_\pm$ from \eqref{eq:neweigenvalues} determine the dynamics of the full system. In general the true dynamics will deviate from the na\"{\i}ve single sector evolution. As explained in section \ref{sec:nonstablehiddensector} it is necessary to put constraints on the full system for the true dynamics to still (largely) agree with the initial na\"{\i}ve dynamics. We will quantify these constraints in terms of the hidden sector light mode $\lambda_q$ and the dynamical cross coupling $|A_{11}|$ between sectors. The results are graphically summarized in figures \ref{fig:etaplus} and \ref{fig:etamin}. In section \ref{sec:sugradata} and figure \ref{fig:etasugra} we will discuss the result again but then interpreted from the viewpoint of supergravity. Finally we will explain that a simple application of these bounds implies that the Standard Model cannot be ignored during cosmological inflation, if Standard Model supersymmetry breaking is independent of the inflaton sector.

\subsection{Conditions on the hidden sector data}
From \eqref{eq:neweigenvalues} we see that the light modes $\lambda_\phi,\lambda_q$ from the two separate sectors mix through a cross coupling $|A_{11}|$ and combine to the true eigenvalues $\mu_\pm$ of the full two-sector system. As explained in \ref{sec:nonstablehiddensector}, for the inflaton sector to still describe the cosmological evolution and the $\eta$-parameter reliably, the three constraints it must obey are (1) the bound arising from demanding a hierarchy between $\mu_\pm$ to prevent multifield effects, (2) the bound arising from demanding the second slow-roll parameter $\mu_-=\eta$ to not change its value too much and (3) the bound from demanding that $\eta$ is mostly determined by the $\phi$-sector rather than the $q$-sector.

To prevent multifield effects from setting in we take as a minimum hierarchy that $\mu_+$ is at least five times as heavy as $\mu_-$ in units of the scale of the problem, $|\mu_{-}|$, \begin{equation}\label{eq:hierarchy}
 \frac{\mu_+-\mu_-}{|\mu_-|}>5.
\end{equation} This bound is rather arbitrary, but clearly a hierarchy between $\mu_{+}$ and $\mu_{-}$ must exist. Calculations in \cite{Peterson:2010np} show that for a mass hierarchy $\lesssim5$ multifield effects are typically important.

The second bound is given by the $A_{11}$-dependence of $\mu_-$. The value of the second slow-roll parameter from the single field inflationary model only is $\etanaive=\lambda_\phi$. In the full two-sector system, $\mu_-$ takes over the role as the true second slow-roll parameter $\eta_{\mathrm{true}}=\mu_-$. The contribution to the actual $\eta$-parameter from the presence of the hidden sector is therefore
\begin{equation}
\Delta\eta=\mu_--\lambda_\phi=\frac{1}{2}\left[(\lambda_q-\lambda_\phi)-\sqrt{(\lambda_q-\lambda_\phi)^2+4|A_{11}|^2}\right] \;,
\end{equation}
which is always negative. We argue that this difference should stay within $|\Delta\eta/\lambda_\phi|< 0.1$, i.e. $\eta$ should not change by more than $10\%$. This choice for the range of $\eta$ is given by current experimental accuracy. Current experiments can only determine $n_{s}=1-6\epsilon+2\eta$. WMAP has a $1\sigma$ error of $6.53\%$ \cite{Komatsu:2010fb}, Planck will have an error of $0.70\%$ \cite{:2006uk}. For $n_{s}-1$, assuming $0.96$, this gives a $17.5\%$ error on the combination of $- 6\epsilon +2\eta$, which means an uncertainty of about $10\%$ on the value of $\eta$.

We will examine $\lambda_{q}, A_{11}$ in units of $|\lambda_\phi|$ and exclude regions in which the hidden sector affects the tuned inflationary sector too much. The analysis is best done separately for the cases $\lambda_\phi=\etanaive>0$ and $\lambda_\phi=\etanaive<0$ because of the qualitative differences between these cases.

\subsubsection{The case $\etanaive>0$}\label{sec:etalargerzero}
\FIGURE[tb]{
\includegraphics[width=0.6\textwidth]{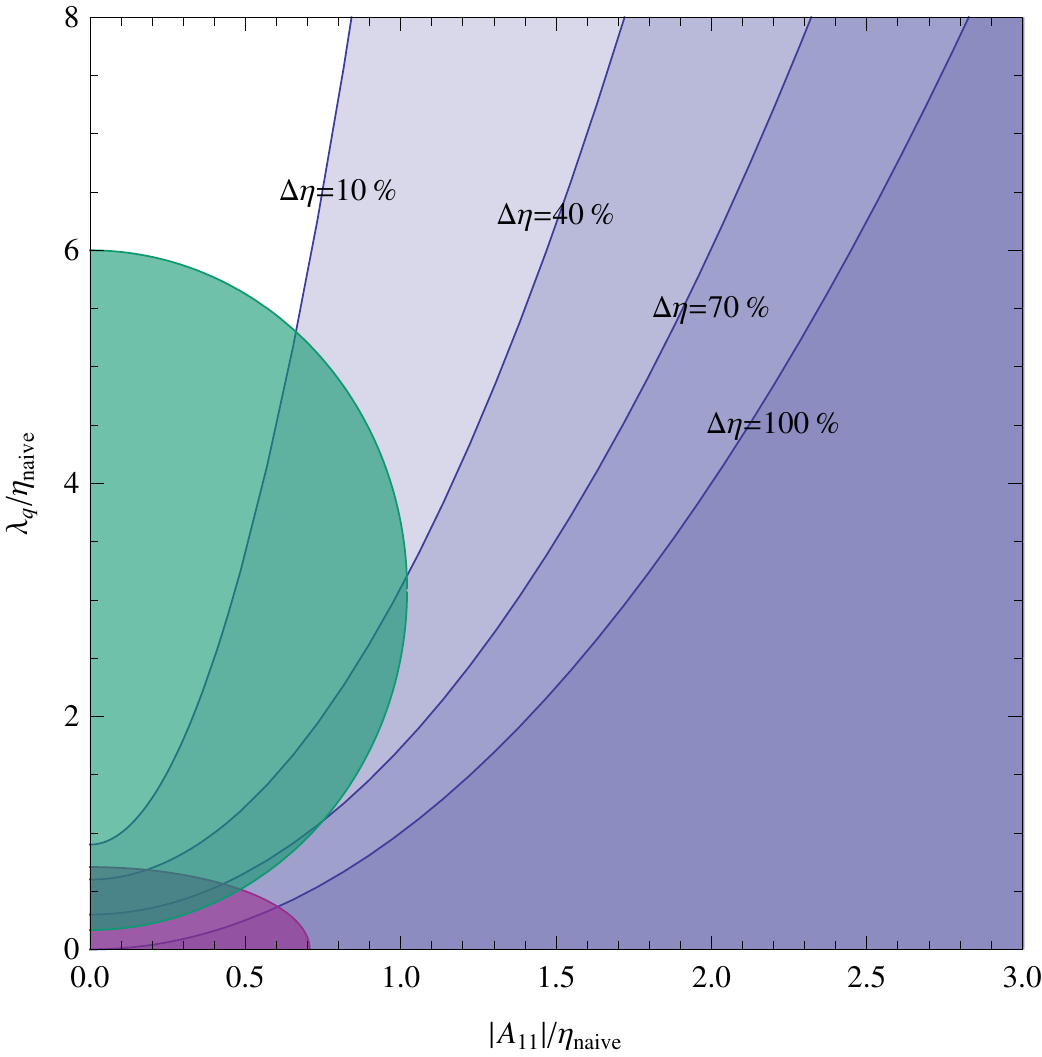}
\caption{Bounds from a dynamical hidden sector for $\etanaive>0$. The multifield constraint excludes an ellipse near the $\lambda_q$-axis (shaded in green). The bound from having too much effect on $\eta$ excludes large $|A_{11}|$ (shaded with increasing intensities of blue for larger deviations). Around $\lambda_q=A_{11}=0$ the hidden sector mode $\lambda_q$ rather than $\lambda_\phi$ determines $\eta$, excluding that region as well (shaded in purple).}\label{fig:etaplus} }

We first examine the hierarchy bound as explained above and focus first on the situation where $\mu_->0$. In this case \eqref{eq:hierarchy} means that we demand
\begin{equation}
\frac{\mu_+-6\mu_-}{\lambda_\phi}=\frac{1}{2}\left[-5\left(\frac{\lambda_q}{\lambda_\phi}+1\right)+7\sqrt{\left(\frac{\lambda_q}{\lambda_\phi}-1\right)^2+4\left(\frac{|A_{11}|}{\lambda_\phi}\right)^2}\right]>0,
\end{equation}
which allows us to solve $\lambda_q/\lambda_\phi$ as a function of $|A_{11}|/\lambda_\phi$,
\begin{equation}
\left(\frac{12}{35}\right)^2\left(\frac{\lambda_q}{\lambda_\phi}-\frac{37}{12}\right)^2+\left(\frac{2\sqrt{6}}{5}\right)^2\left(\frac{|A_{11}|}{\lambda_\phi}\right)^2=1.
\end{equation}
This excludes everything inside the ellipse demarcating the green region in figure \ref{fig:etaplus}. The case $\mu_-<0$ is not relevant as it is already excluded by the second bound.

For this second bound, to be somewhat more general than the observationally inspired constraint $\Delta\eta/\lambda_\phi>-0.1$, we give the bound $\Delta\eta/\lambda_\phi>-f$. Solving for $\lambda_q$ this gives
\begin{equation}
\frac{\lambda_q}{\lambda_\phi}>1-f+\frac{1}{f}\left(\frac{|A_{11}|}{\lambda_\phi}\right)^2,
\end{equation}
as is indicated in blue in figure \ref{fig:etaplus}. Note that since the true value of $\eta$ is always lower than $\etanaive$ (see \cite{Dong:2010in} for some specific examples), a change in $\eta$ of $100\%$ means that $\eta$ changes sign from its na\"{\i}ve value. This shows that we were justified to only consider positive $\mu_-$ in the hierarchy bound earlier.

The third bound is given by a $\lambda_q$-dominance in $\mu_-$. Since $\lambda_\phi$ and $\lambda_q$ are treated on equal footing in $\mu_-$, the true $\eta$ is dominantly determined by the smallest eigenvalue, which is not necessarily $\lambda_\phi$. When $\lambda_\phi\gg\lambda_q$ and $\lambda_\phi\gg |A_{11}|$ we see immediately that the true $\eta=\mu_-$ is determined by $\lambda_q$ and is \emph{independent} of $\lambda_\phi$,
\begin{equation}
\mu_-=\frac{1}{2}\left[(\lambda_q+\lambda_\phi)-\lambda_\phi\left(1-\frac{\lambda_q}{\lambda_\phi}+\mathcal{O}\left(\frac{\lambda_q^2}{\lambda_\phi^2},\frac{|A_{11}|^2}{\lambda_\phi^2}\right)\right)\right].
\end{equation}
It is clear that this arguments excludes the lower left corner of parameter space. We will take the bound to be $1/\sqrt{2}$ such that $\left(\lambda_q/\lambda_\phi\right)^2,\left(|A_{11}|/\lambda_\phi\right)^2<1/2\ll 1$, the radius of convergence of this Taylor expansion. Contrarily to the somewhat debatable bounds imposed by $\Delta\eta/\lambda_\phi$, the points within this circle are truly excluded because they violate one of the core assumptions in the approach, viz.~that the $\phi$-sector is responsible for all cosmological dynamics including determining the value of $\eta$. The circle
\begin{equation}
\left(\frac{\lambda_q}{\lambda_\phi}\right)^2+\left(\frac{|A_{11}|}{\lambda_\phi}\right)^2=\frac{1}{2},
\end{equation}
is indicated as the purple region in the figure.

In figure \ref{fig:etaplus} we have indicated in which regions of $\lambda_q/\lambda_\phi$- and $|A_{11}|/\lambda_\phi$-parameter space the effects of a hidden sector can be rightfully ignored. We have shown that all negative values of $\lambda_q$ are excluded and only in the region with large $\lambda_q/\lambda_\phi$ and small $|A_{11}|/\lambda_\phi$ there are no large effects from the hidden sector. This result is qualitatively easily understood, as the hidden sector with broken supersymmetry will still decouple if the masses in the hidden sector are truly large. We argue that this possibility is too easily assumed to be the case in the literature without considering the actual hidden constraints it imposes on the hidden sector. These hidden assumptions should be mentioned explicitly and one should show that they can be obtained.

\subsubsection{The case $\etanaive<0$}
\FIGURE[tb]{
\includegraphics[width=0.6\textwidth]{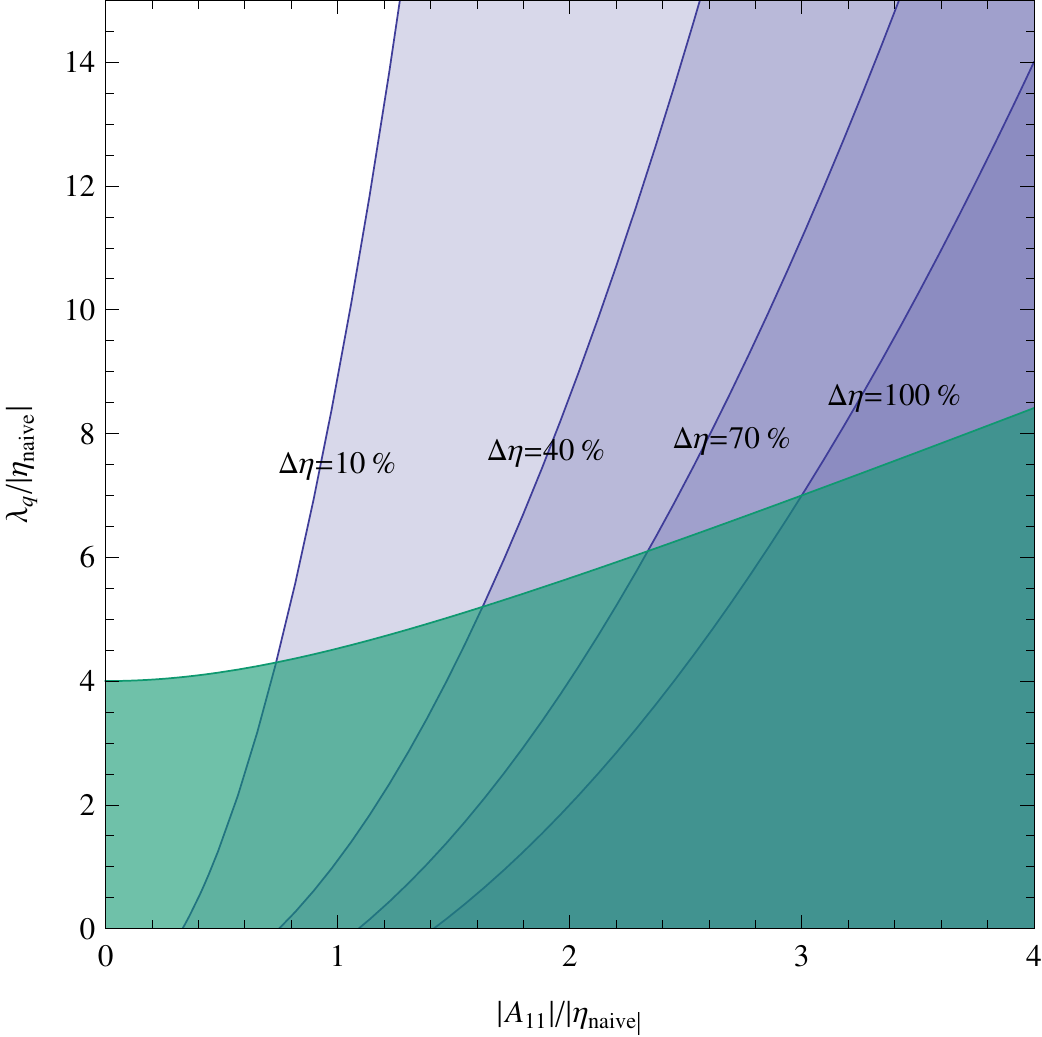}
\caption{Bounds from a dynamical hidden sector for $\etanaive<0$. The multifield bound excludes a hyperbola starting at $\lambda_q=4|\lambda_\phi|$ and, in particularly small $\lambda_q$ (shaded in green). The bound from having too much effect on $\eta$ excludes the large $|A_{11}|$-region (shaded with increasing intensities of blue for larger deviations), but leaves open in particular the full range of $\lambda_q$.}\label{fig:etamin} }

In the case that $\lambda_\phi=\etanaive$ is negative, the last bound of section \ref{sec:etalargerzero} does not impose any condition on $\lambda_q/|\lambda_\phi|,|A_{11}|/|\lambda_\phi|$-parameter space. When $\lambda_\phi<0$, i.e. when $\lambda_\phi=-|\lambda_\phi|$, the eigenvalues can be written as
\begin{equation}
 \mu_\pm=\frac{|\lambda_\phi|}{2}\left[\left(\frac{\lambda_q}{|\lambda_\phi|}-1\right)\pm\sqrt{\left(\frac{\lambda_q}{|\lambda_\phi|}+1\right)^2+4\left|\frac{A_{11}}{\lambda_\phi}\right|^2}\right]\;,
\end{equation}
which means that $\mu_-$ is not determined by $\lambda_q$ to first order in $\lambda_q/|\lambda_\phi|$ but by $\lambda_\phi$ as should be,
\begin{equation}
 \mu_-=\frac{|\lambda_\phi|}{2}\left[\left(\frac{\lambda_q}{|\lambda_\phi|}-1\right)-\left(1+\frac{\lambda_q}{|\lambda_\phi|}+\ldots\right)\right]\;.
\end{equation}
However, by the hierarchy bound the small $\lambda_q/|\lambda_\phi|$-regime does get excluded. Since $\mu_-$ is always negative in this case,
\begin{equation}
\mu_-\leq \frac{|\lambda_\phi|}{2}\left[\left(\frac{\lambda_q}{|\lambda_\phi|}-1\right)-\left|\frac{\lambda_q}{|\lambda_\phi|}+1\right|\right]=-|\lambda_\phi|\;,
\end{equation}
equation \eqref{eq:hierarchy} translates into
\begin{equation}
\frac{\mu_++4\mu_-}{|\lambda_\phi|}=\frac{1}{2}\left[5\left(\frac{\lambda_q}{|\lambda_\phi|}-1\right)-3\sqrt{\left(\frac{\lambda_q}{|\lambda_\phi|}+1\right)^2+4\left|\frac{A_{11}}{\lambda_\phi}\right|^2}\right]>0\;.
\end{equation}
This excludes everything beneath the upper branch of the hyperbola given by the line
\begin{equation}
\frac{\lambda_q}{|\lambda_\phi|}>\frac{17}{8}+\frac{1}{8}\sqrt{15^2+28\left|\frac{A_{11}}{\lambda_\phi}\right|^2}\;,
\end{equation}
which is shaded green region in figure \ref{fig:etamin}.

The final constraint on the parameter space comes from the bound on the change in $\eta$, see the previous paragraph on the $\etanaive>0$-case for a discussion. In the blue region in figure \ref{fig:etamin} we have indicated the bound $|\Delta\eta/\lambda_\phi|<f$, which means
\begin{equation}
\frac{\lambda_q}{|\lambda_\phi|}>-f+\left|\frac{A_{11}}{\lambda_\phi}\right|^2\;,
\end{equation}
for different fractions of $f$.

In figure \ref{fig:etamin} we have indicated in which regions of $\lambda_q/|\lambda_\phi|$- and $|A_{11}|/|\lambda_\phi|$-parameter space the effects of a hidden sector can be rightfully ignored after imposing both constraints. As in the case for $\etanaive>0$, the only allowed region is for large $\lambda_q/|\lambda_\phi|$ and small $|A_{11}|/|\lambda_\phi|$. Note that all values of $\lambda_q<4$ are explicitly excluded by the imposed bounds.

\subsection{Conditions on supergravity models\label{sec:sugradata}}
\FIGURE[tb]{
\includegraphics[width=0.45\textwidth]{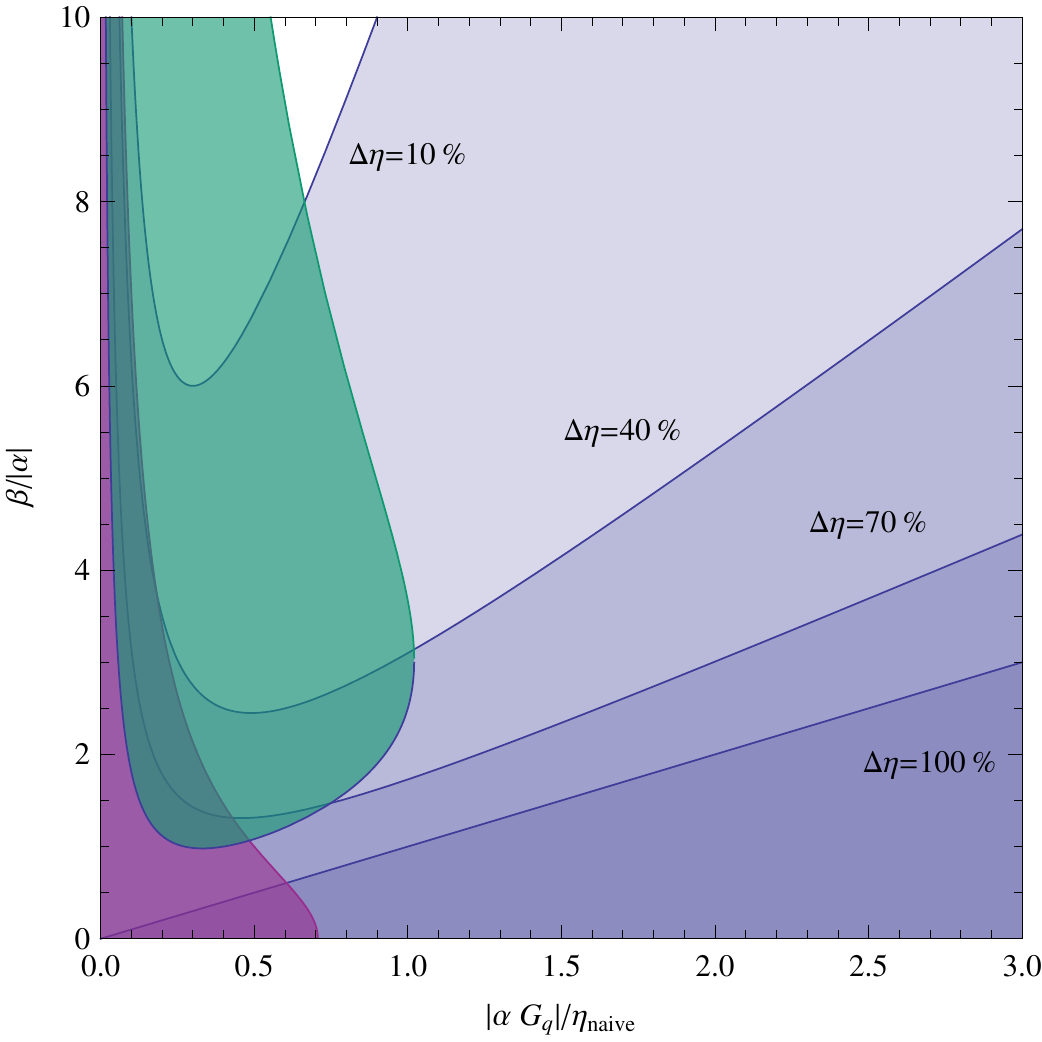}\hfill\includegraphics[width=0.45\textwidth]{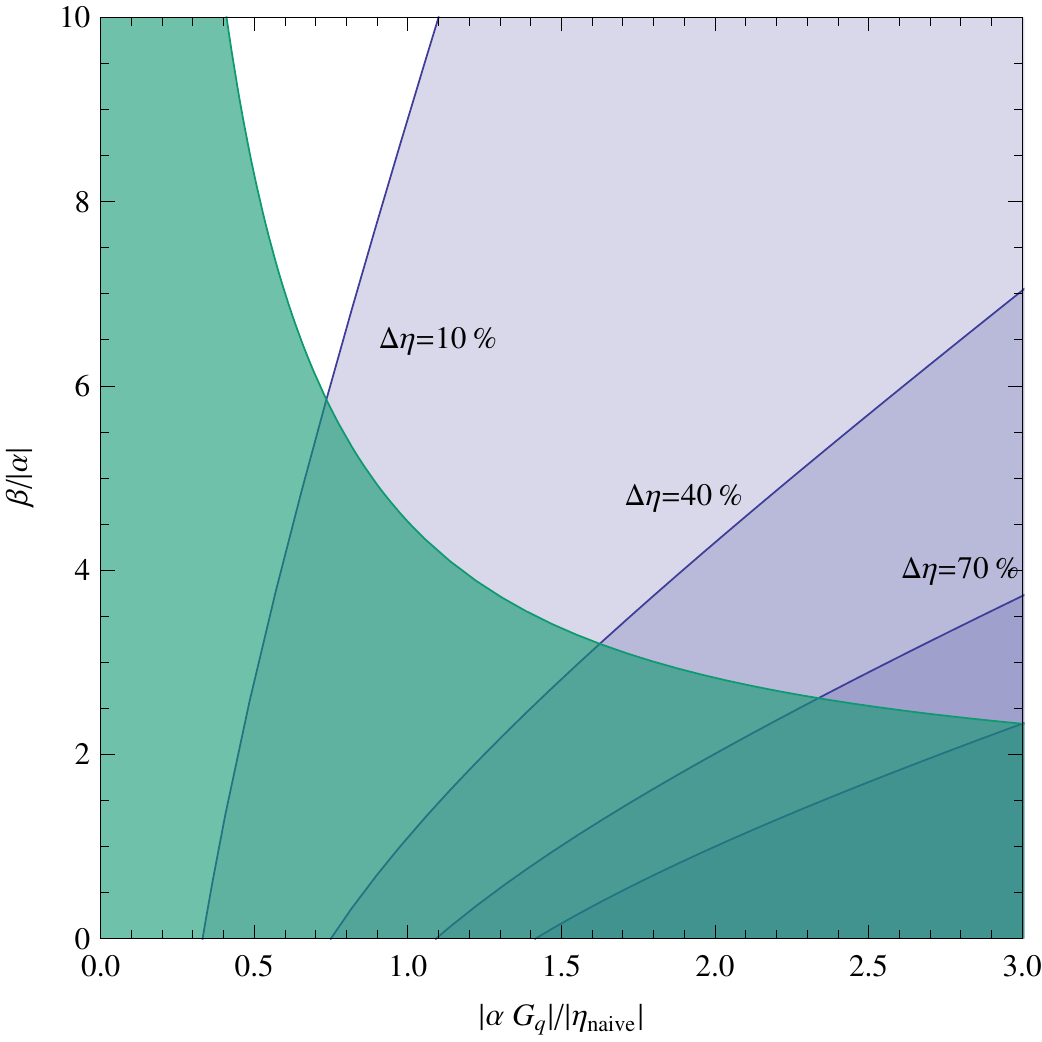}
\caption{Excluded regions for the supergravity parameter range for $|G_q|$ and $\beta$, which contains in particular $\nabla_q\nabla_q G_q $, in units of $|\etanaive|$ and $|\alpha|$, which contains $\epsilon_\phi$ and $G_\phi$. The indicated regions come from the multifield bound (shaded in green), the correct identification of sectors (shaded in purple) and allowing only for small deviations of $\eta$ (shaded in higher intensities of blue for larger deviations). The left (right) picture describes the case $\etanaive>0$ ($\etanaive<0$).}\label{fig:etasugra}}

In principle, figures \ref{fig:etaplus} and \ref{fig:etamin} provide all the information needed to verify whether the hidden sector of a given model may be neglected while studying the inflationary dynamics. Through equations (\ref{eq:lambdaq}--\ref{eq:Amix}) and the expressions for $V_{\alpha\beta}$ as summarized in appendix \ref{app:sugrarelations}, one can explicitly calculate the corresponding $\lambda_q$ and $A_{11}$ for a given model and compare them with the figures. However, we would like to have some direct intuition about the dependence of the excluded regions on the supergravity data. In this section we will investigate how much we can say about this in general without having to specify a model. The main question to answer is whether the fact that $\lambda_q$ and $A_{11}$ are determined by a supergravity theory, provides any additional constraint on which regions are obtainable to begin with. The answer to this question turns out to be that a priori supergravity is not restrictive enough to exclude any of the regions in $\lambda_q,A_{11}$-parameter space.

The easiest way to translate figures \ref{fig:etaplus} and \ref{fig:etamin} in terms of supergravity data would be to simply map the regions into supergravity parameter space. Unfortunately the expressions \eqref{eq:lambdaq} and \eqref{eq:Amix} are highly nonlinear and depend on too many supergravity variables to conveniently represent figures \ref{fig:etaplus} and \ref{fig:etamin} in terms of supergravity data. However, for small $|G_q|$ this does turn out to be possible.

Using the expressions for $V_{\alpha\beta}$ in \eqref{eq:Amix}, yields
\begin{align}\label{eq:Asugradata}
A_{11}&=\alpha(\phi,\bar{\phi},q,\bar{q})|G_q|,\qquad\textrm{with}\\
\alpha(\phi,\bar{\phi},q,\bar{q})&=\frac{G^{\phi\bar{\phi}}}{2}\left(\widehat{G_{\bar{q}}}-\widehat{V_{\bar{q}\bar{q}}}\widehat{G_q}\right)\left(\left(\frac{V_\phi}{V}-G_\phi\right)-\widehat{V_{\phi\phi}}\left(\frac{V_{\bar{\phi}}}{V}-G_{\bar{\phi}}\right)\right).\nonumber
\end{align}
From this equation we learn that $A_{11}$ vanishes in the limit $G_q\to 0$, which makes sense as we know that the two sectors should decouple in the limit of restored supersymmetry. It is difficult to retrieve more information from this explicit expression of $A_{11}$ in terms of supergravity data. In principle $A_{11}(|G_q|,\ldots)$ may be inverted to give some function $|G_q|(A_{11},\ldots)$, but this is more tricky than \eqref{eq:Asugradata} suggests. Although we have managed to extract one factor of $G_q$, the function $\alpha(\phi,\bar{\phi},q,\bar{q})$ still depends on $G_q$ through the phases of $\widehat{V_{\bar{q}\bar{q}}}$ and $\widehat{V_{\phi\phi}}$, making it hard to perform the inversion explicitly.

The expression for $\lambda_q$ looks even worse,
\begin{equation}\label{eq:lambdasugradata}
\lambda_q=\frac{G^{q\bar{q}}}{V}\left(V_{q\bar{q}}-\sqrt{V_{qq}V_{\bar{q}\bar{q}}}\right).
\end{equation}
At this stage we have even refrained from substituting in the expressions for $V_{q\bar{q}}$, $V_{qq}$ and its complex conjugate. The square root clearly shows that the dependence of $\lambda_q$ on $|G_q|$ and the other supergravity data is extremely involved and difficult to invert. To get a useful expression we revert to the result of section \ref{sec:zeromode} and consider $\lambda_q$ in the small $|G_q|$-regime by performing a Taylor expansion. Copying from \eqref{eq:zeromassmode}, we find
\begin{align}\label{eq:lambdaGq}
\lambda_q &=\beta(\phi,\bar{\phi},q,\bar{q})|G^q|+\mathcal{O}(|G_q|^2),\qquad\textrm{with}\\
\beta(\phi,\bar{\phi},q,\bar{q})&=\frac{G^{q\bar{q}}}{e^{-G}V}\textrm{Re}\big\{(\nabla_q\nabla_qG_q)\widehat{G^q}^3\big\}.\nonumber
\end{align}

Having obtained the relations \eqref{eq:Asugradata} and \eqref{eq:lambdaGq} we can now accommodate the reader with a graph of the allowed and excluded regions directly in terms of the supergravity data. For small $G_q\ll 1$ both $\lambda_q$ and $|A_{11}|$ scale linearly with $G_q$, making it relatively easy to rewrite the bounds we found  $\lambda_q/|\lambda_\phi|=\lambda_q/|\lambda_\phi|\left(|A_{11}|/|\lambda_\phi|\right)$ in terms of $G_q$, $\alpha$ and $\beta$ as $\beta/|\alpha|=\beta/|\alpha|\left(|\alpha G_q|/|\lambda_\phi|\right)$. The resulting figure is depicted in \ref{fig:etasugra}. Note that $\alpha$ and $\beta$ are still underdetermined --- depending on $R_{q\bar{q}q\bar{q}}$ and $\nabla_{q}\nabla_{q}G_{q}$ at higher orders in $|G_{q}|$ --- and are naturally of order $1$. It is these numbers that determine where in figure \ref{fig:etasugra} the model under investigation lies.

\subsection{Inflation and the Standard Model\label{sec:inflationSM}}
\FIGURE[ht]{
\includegraphics[width=0.45\textwidth]{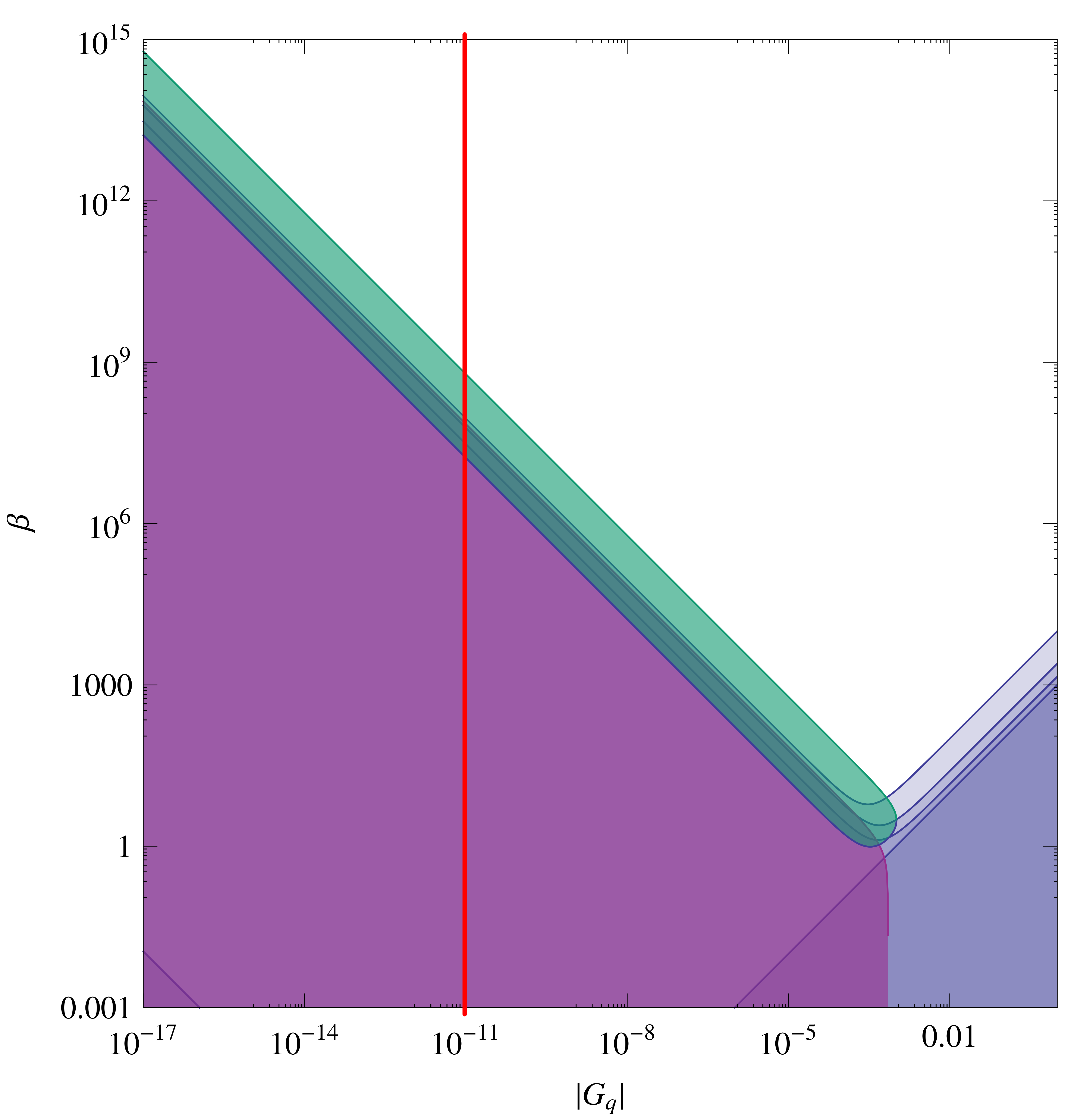}\hfill\includegraphics[width=0.45\textwidth]{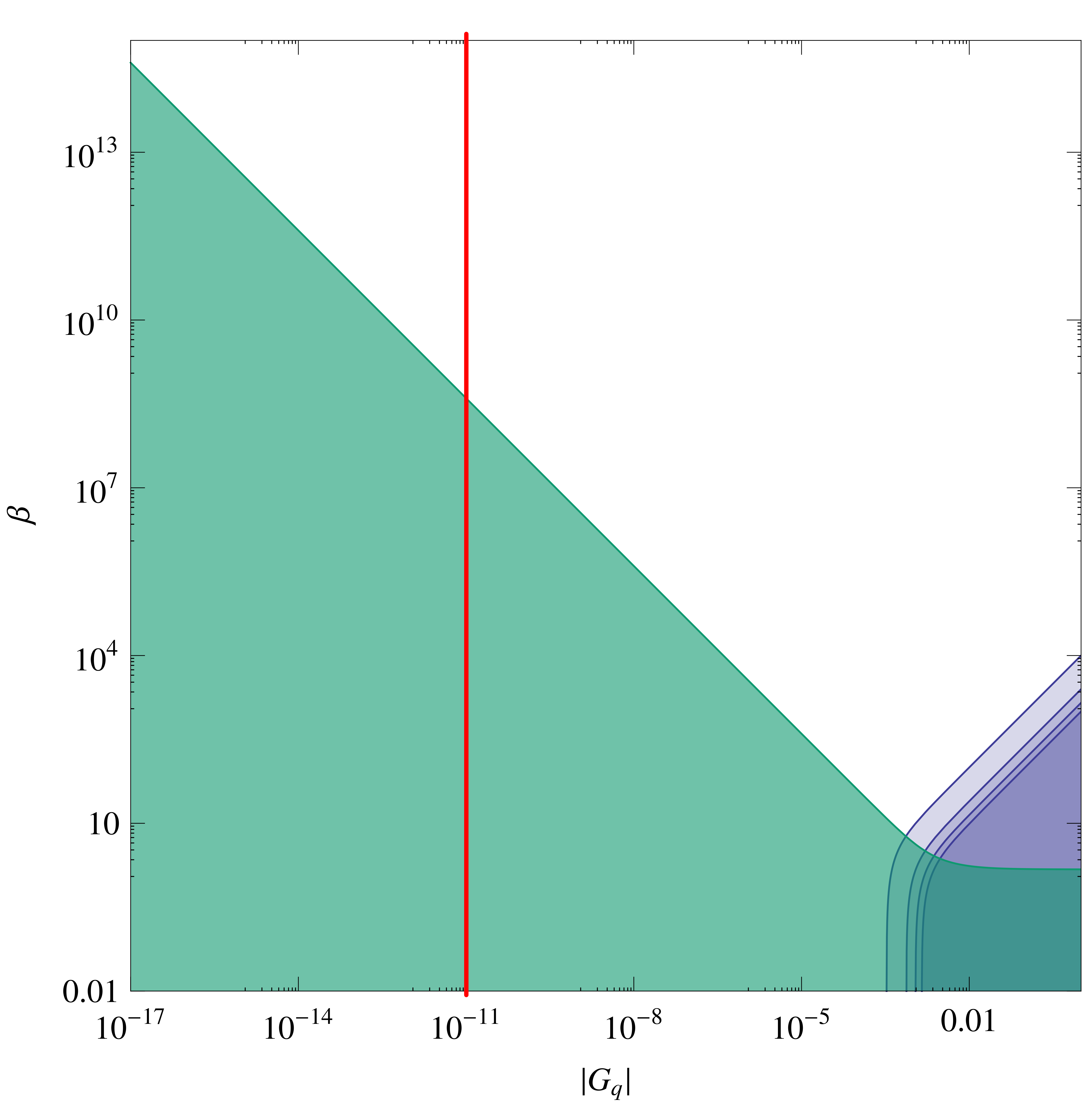}
\caption{The effects of the multifield bound (shaded in green), the identification of the correct inflaton sector (shaded in purple) and the small deviations of $\eta$ (shaded in blue) on a doubly logarithmic scale for $\etanaive>0$ (left) and $\etanaive<0$ (right). The approximate location of the Standard Model supergravity data is indicated with a red bar, showing that a large range of parameters is excluded. In this plot $\alpha=1$ and $\lambda_\phi=\etanaive=10^{-3}$.}\label{fig:loglogeta}}
As a simple application of the previous section, we can consider to what extent the Standard Model ought to be included in any reliable supergravity model for cosmological inflation. Our current understanding of Nature includes a present-day supersymmetrically broken Standard Model after an inflationary evolution right after the big bang. As such the combined model is exactly that of a two-sector supergravity theory with an inflationary and a hidden sector whose ground state breaks supersymmetry in which it resides throughout the inflationary era.

Supersymmetry in the Standard Model sector can either have been broken by gravity mediation of the inflaton sector or by a mechanism in the Standard Model sector itself. The first situation would be consistent approach as far as our analysis goes: as $G_q=0$ the sector decouples from the inflationary dynamics, can be stabilized and the slow-roll parameters are reliably determined from the inflaton sector alone. Nevertheless, from the point of view of our understanding of the Standard Model it would be unsatisfactory to not know the precise mechanism behind its supersymmetry breaking and (complete) models describing such mechanisms would still have to be analysed to shed light on the situation.

In the second situation $G_q\neq 0$ and we should apply the results of the previous sections. The field $q$ may be seen as some light scalar degree of freedom in the (supersymmetrically broken) Standard Model. We assume the standard lore, that supersymmetry is broken in the Standard Model at a scale of about 1 $\TeV$. In the $F$-term scalar potential, this scale enters via $G_q$. To determine the correct numerical value, we relate our dimensionless definition of the K\"{a}hler function to the standard dimensionful definition. Dimensionful quantities are denoted with a tilde in the following.\footnote{E.g. in dimensionful units $[\widetilde{G}]=\textrm{mass}^2$ and $[\widetilde{q}]=\textrm{mass}$, while our conventions are $[G]=[q]=0$. To relate $G_q$ to $\widetilde{G}_{\widetilde{q}}$ we can use the expression $\left[G_q\right]=\frac{[\widetilde{G}_{\widetilde{q}}]}{\mpl}$.} We recall from section \ref{sec:canonicalcoupling} that in order to have a non-vanishing vacuum energy, the superpotential in both sectors must have a non-zero constant term $W^{(1)}_0=m_{\Lambda}^{(1)}/\mpl$, $W^{(2)}_0 =m_{\Lambda}^{(2)}/\mpl$, which accounts for the always present gravitational coupling between the sectors. Hence, the dimensionful constant term in the total superpotential \eqref{eq:totalsplitpotential} has value $\widetilde{W}^{\textrm{tot}}_{0}=W_{0}^{(1)}W_{0}^{(2)}\mpl^{3}=m_{\Lambda}^{(1)}m_{\Lambda}^{(2)}\mpl$. In contrast, the supergravity quantities $\widetilde{K}^{(2)}$ and $\widetilde{W}^{(2)}_{\mathrm{eff}} = \widetilde{W}_{0}^{(1)}\widetilde{W}_{\textrm{global}}^{(2)}$ describing the Standard Model are naturally of the order of the $\textrm{TeV}$-scale, $[\widetilde{W}^{(2)}_{\mathrm{eff}}]=\textrm{TeV}^3$,  $[\partial_{\tilde{q}}\widetilde{K}^{(2)}]=\textrm{TeV}$. We relate the scale of supersymmetry breaking $\widetilde{G}_{\widetilde{q}}$ to the superpotential via
\begin{equation}
 \widetilde{G}_{\tilde{q}} = \frac{\mpl^2}{\widetilde{W}}\left(\partial_{\tilde{q}} \widetilde{W} + \frac{\partial_{\tilde{q}} \widetilde{K}^{(2)}}{\mpl^2}\widetilde{W}\right) \;,
\end{equation}
which is naturally of order
\begin{equation}
\left[\widetilde{G}_{\tilde{q}}\right]=
\frac{\mpl^2}{m_{\Lambda}^{(1)}m_{\Lambda}^{(2)}\mpl + \ldots}\left(\textrm{TeV}^2 +
\frac{\textrm{TeV}}{\mpl^2}(m_{\Lambda}^{(1)}m_{\Lambda}^{(2)}\mpl + \ldots)\right)=\frac{\mpl \textrm{TeV}^{2}}{m_{\Lambda}^{(1)}m_{\Lambda}^{(2)}} + \textrm{TeV} + \ldots\;,
\end{equation}
where the $\ldots$ are of subleading order. We expect that $m_{\Lambda}^{(1)}$, the constant term of the inflaton sector, is of order $[H] = 10^{-5}\mpl$, while $[m_{\Lambda}^{(2)}] = \textrm{TeV}$. Hence, translating back to dimensionless units, we find $G_q \sim 10^{-11}$.

Taking the kinetic gauge, i.e. a canonical K\"{a}hler metric $G_{\phi\bar{\phi}}=1$, we can easily find the natural value of $\alpha$. From \eqref{eq:Asugradata} we see that $\alpha$ depends on $\eps_\phi$ and $G_\phi$ via
\begin{equation}
\alpha\propto \sqrt{\eps_\phi}-G_\phi,
\end{equation}
modulo some unknown but negligible phase factors. $G_\phi$ is of order $\sqrt{3}$ in order to have a potential $V>0$. Since $\eps_\phi$ is of order $\mathcal{O}(10^{-3})$, the value of $|\alpha|$ is of order unity. For a rough estimate for $\etanaive\sim 10^{-3}$ we can therefore pinpoint the Standard Model as indicated in figure \ref{fig:loglogeta}. In both cases, $\etanaive>0$ as well as $\etanaive<0$, the lightest supersymmetric particle is too light for the single sector inflationary dynamics to truly describe the full system. Any tuned and working inflationary supergravity model in which the Standard Model is assumed to not take part considerably in the cosmic evolution, requires implicit assumptions on the Standard Model that either the inflaton sector is responsible for Standard Model supersymmetry breaking through gravity mediation or the masses of its scalar multiplets are unnaturally large in terms of the now independent Standard Model supersymmetry breaking scale.

\section{Conclusions}
In this paper we have studied the effect of hidden sectors on the finetuning of $F$-term inflation in supergravity, identifying a number of issues in the current methodology of finetuning inflation in supergravity. Finetuning inflationary models is only valid when the neglected physics does not affect this finetuning, in which case the inflationary physics can be studied independently. As shown in figures \ref{fig:etaplus} and \ref{fig:etamin} this assumption holds only under very special circumstances. The reason is that the everpresent gravitational couplings will always lead to a mixing of the hidden sectors with the inflationary sector, even in the case of the most minimally coupled action \eqref{eq:splitaction}. For a hidden sector vacuum that preserves supersymmetry, the sectors decouple consistently \cite{Choi:2004sx,deAlwis:2005tf,deAlwis:2005tg,Achucarro:2007qa,Achucarro:2008sy}. However, for a supersymmetry breaking vacuum the inflationary dynamics is generically altered, where the nature and the size of the change depends on the scale of supersymmetry breaking.

For a hidden sector with a low scale of supersymmetry breaking, like the Standard Model, the cross coupling scales with the scale of supersymmetry breaking, and is therefore typically small. Yet, as shown in section \ref{sec:zeromode}, also the lightest mass of the hidden sector scales with the scale of supersymmetry breaking within that sector. This light mode is strongly affected by the inflationary physics and thus evolves during inflation. Therefore, any single field analysis is completely spoiled as discussed in section \ref{sec:inflationSM}.

For massive hidden sectors, the problem is more traditional. For a small hidden sector supersymmetry breaking scale, one has a conventional decoupling as long as the lightest mass of the hidden sector is much larger than the inflaton mass. However, for large hidden sector supersymmetry breaking, this intuition fails. Then, the off-diagonal terms in the mass matrix \eqref{eq:massmatrix} will lead to a large correction of the $\eta$-parameter.

To conclude, any theory that is working by only tuning the inflaton sector has made severe hidden assumptions about the hidden sector, which typically will not be easily met. Methodologically the only sensible approach is to search for inflation in a full theory, including knowledge of all hidden sectors.

\acknowledgments
We thank A. Ach\'ucarro, B. Underwood, and B.J. van Tent for discussions, as well as G. Shiu who participated at early stages of this project. GAP wishes to thank the Lorentz Institute (Leiden) for their hospitality during the preparation of the manuscript. KS thanks the Galileo Galilei Institute for Theoretical Physics for the hospitality and the INFN for partial support during the completion of this work. This research was supported in part by Conicy under a Fondecyt Initiation on Research Grant 11090279 (G. A. Palma), a VIDI Innovative Research Incentive Grant (K. Schalm) from the Netherlands Organisation for Scientific Research (NWO), a VICI Award (A. Ach\'ucarro) from the Netherlands Organisation for Scientific Research (NWO) and the Dutch Foundation for Fundamental Research on Matter (FOM).

\appendix
\section{Some supergravity relations}
\label{app:sugrarelations}
For easy reference to the reader, we use this appendix to state the relevant derivatives of the supergravity potential of a two-sector system coupled via
\begin{equation}\label{eq:appslitG}
G(\phi^i,\bar{\phi}^{\bar{\imath}},q^a,\bar{q}^{\bar{a}}) =
G^{(1)}(\phi^i,\bar{\phi}^{\bar{\imath}}) +
G^{(2)}(q^a,\bar{q}^{\bar{a}})\;.
\end{equation}
We use middle-alphabet Latin indices $\{i,\bar{\imath}\}$ to denote the fields in the inflationary sector, beginning-alphabet Latin indices $\{a,\bar{a}\}$ to denote the fields in the hidden sector and Greek indices $\{\alpha,\bar{\alpha}\}$ to denote the full system. Derivatives with respect to these fields are denoted by subscripts, e.g. $\partial_iG=G_i$ and $\partial_i\partial_jG=G_{ij}$. The Hessian $G_{\alpha\bar{\beta}}$ describes the metric of the (product-) manifold parametrised by the fields. This is a K\"ahler manifold and hence $\nabla_\alpha G_{\bar{\beta}}=G_{\alpha\bar{\beta}}$.

The supergravity potential is
\begin{equation}
V = e^G(G_\alpha
G^\alpha-3)=e^G(G_{\bar{\alpha}}G^{\bar{\alpha}}-3)=e^{G}(G_{a}G^{a}
+ G_{i}G^{i} - 3)\;.
\end{equation}
Its covariant derivatives are denoted with subscripts (note that this is a different convention than the one used for the K\"ahler function $G$), e.g. $\nabla_iV=\partial_iV=V_i$ and $\nabla_i\nabla_j V=V_{ij}$. In terms of derivatives of $G$, the first derivatives of $V$ are given by
\begin{align}
\label{eq:Vi}
V_{i}&= G_{i}V + e^{G}\left( (\nabla_{i}G_{j})G^{j} + G_{i}\right)\;,\\
\label{eq:Vqbar}
V_{\bar{\imath}} &=
G_{\bar{\imath}}V+e^G\left((\nabla_{\bar{\imath}}G_{\bar{\jmath}})G^{\bar{\jmath}}+G_{\bar{\imath}}\right)\;,
\end{align}
and similar expressions for $V_a$ and
$V_{\bar{a}}$. The Hessian of covariant derivatives is
\begin{align}\label{eq:Vij}
 V_{ij}&=\nabla_{i} G_{j} V+G_{i} V_{j}+G_j V_i-G_{i}G_{j}V+e^{G}\left[(\nabla_{i}\nabla_{j} G_{k})G^{k}+2\nabla_{i} G_{j}\right]\;,\\
 \label{eq:Vijbar}
  V_{i\bar{\jmath}}&=G_{i\bar{\jmath}}V+G_{i} V_{\bar{\jmath}}+G_{\bar{\jmath}}V_{i}-G_{i} G_{\bar{\jmath}}V +e^{G}\left[R_{i\bar{\jmath}k\bar{l}}G^{k}G^{\bar{l}}+G^{k\bar{l}}\nabla_{i} G_{k}\nabla_{\bar{\jmath}}G_{\bar{l}}+G_{i\bar{\jmath}}\right]\;,\\
  \label{eq:Via}
  V_{i a}&=\nabla_{a} G_{i} V+G_{i} V_{a}+G_{a}V_{i}-G_{i} G_{a}V+e^{G}\left[(\nabla_{a}\nabla_{i} G_{\alpha})G^{\alpha}+\nabla_{i} G_{a}+\nabla_{a}G_{i}\right]\nonumber\\
  &=G_{i} V_{a}+G_{a}V_{i}-G_{i} G_{a}V\;,\\
  \label{eq:Viabar}
  V_{i\bar{a}}&=G_{i\bar{a}}V+G_{i} V_{\bar{a}}+G_{\bar{a}} V_{i}-G_{i} G_{\bar{a}}V+e^{G}\left[R_{\alpha\bar{\beta}i\bar{a}}G^{\alpha}G^{\bar{\beta}}+G^{\alpha\bar{\beta}}\nabla_{i} G_{\alpha}\nabla_{\bar{a}}G_{\bar{\beta}}+G_{i\bar{a}}\right] \nonumber \\
  &=G_{i} V_{\bar{a}}+G_{\bar{a}}V_{i}-G_{i} G_{\bar{a}}V\;,
\end{align}
and similar expressions for the other $V_{\alpha\beta}$. The equalities in \eqref{eq:Via} and \eqref{eq:Viabar} are a result of the specific form of the K\"ahler function \eqref{eq:appslitG}.

\section{Mass eigenmodes in a stabilized sector}
\label{app:masslessmode}
In this appendix we provide some intermediate results in the
calculation of
(\ref{eq:zeromassmode}--\ref{eq:heavymassmode}). Using the
expressions as stated in appendix
\ref{app:sugrarelations}, to first order in $|G_q|$, the
second derivatives of the potential are given by
\begin{align}
 V_{qq}&=e^G\left[(2+e^{-G}V)\nabla_qG_q+(\nabla_q\nabla_qG_q)G^q\right]+\mathcal{O}(|G_q|^2)\;, \label{eq:Vqq1} \\
V_{q\bar{q}}&=e^G\left[G_{q\bar{q}}(1+e^{-G}V)+G^{q\bar{q}}(\nabla_qG_q)(\nabla_{\bar{q}}G_{\bar{q}})\right]+\mathcal{O}(|G_q|^2)\;.
\label{eq:Vqq2}
\end{align}
Using the supersymmetry breaking restriction
\eqref{eq:Vq=0condition} in \eqref{eq:Vqq1} and \eqref{eq:Vqq2},
we find
\begin{align}
V_{qq}&=-e^GG_{q\bar{q}}\left[(2+e^{-G}V)(1+e^{-G}V)\widehat{G^q}^{-2}-G^{q\bar{q}}(\nabla_q\nabla_qG_q)G^q\right]+\mathcal{O}(|G_q|^2)\;,\\
V_{q\bar{q}}&=e^G\left[G_{q\bar{q}}(1+e^{-G}V)+(1+e^{-G}V)^2G^{q\bar{q}}G_{q\bar{q}}G_{q\bar{q}}\right]+\mathcal{O}(|G_q|^2)\notag\\
&=e^GG_{q\bar{q}}(2+e^{-G}V)(1+e^{-G}V)+\mathcal{O}(|G_q|^2)\;,
\end{align}
and hence
\begin{align}
|V_{qq}|&=e^GG_{q\bar{q}}(2+e^{-G}V)(1+e^{-G}V)\times\notag\\
&\phantom{=}\times\sqrt{1-\frac{2G^{q\bar{q}}\textrm{Re}\big\{(\nabla_q\nabla_qG_q)G^q\widehat{G^{\bar{q}}}^{-2}\big\}}{(2+e^{-G}V)(1+e^{-G}V)}+\frac{\left|G^{q\bar{q}}(\nabla_q\nabla_qG_q)G^q\right|^2}{(2+e^{-G}V)^2(1+e^{-G}V)^2}}+\mathcal{O}(|G_q|^2)\notag\\
&=e^GG_{q\bar{q}}\left[(2+e^{-G}V)(1+e^{-G}V)-G^{q\bar{q}}\textrm{Re}\big\{(\nabla_q\nabla_qG_q)\widehat{G^q}^3\big\}|G^q|\right]+\mathcal{O}(|G_q|^2)\;.
\end{align}
Then \eqref{eq:massmodes} is evaluated to be
\begin{align}
m_q^-&=e^G G^{q\bar{q}}\textrm{Re}\big\{(\nabla_q\nabla_qG_q)\widehat{G^q}^3\big\}|G^q|+\mathcal{O}(|G_q|^2)\;,\\
m_q^+&=e^G\left[2(2+e^{-G}V)(1+e^{-G}V)-G^{q\bar{q}}\textrm{Re}\big\{(\nabla_q\nabla_qG_q)\widehat{G^q}^3\big\}|G^q|\right]+\mathcal{O}(|G_q|^2)\;.
\end{align}

\providecommand{\href}[2]{#2}\begingroup\raggedright\endgroup

\end{document}